\pdfoutput=1
\documentclass[]{aa}
\usepackage{amsmath}
\usepackage{latexsym}
\usepackage{longtable}
\usepackage{graphics}
\usepackage{epsfig}
\usepackage{graphicx}

\begin{document}

\title{A weakly random Universe?}

\author{V.G.Gurzadyan\inst{1}, A.E.Allahverdyan\inst{1}, T.Ghahramanyan\inst{1},\\ A.L.Kashin\inst{1}, H.G.Khachatryan\inst{1}, A.A.Kocharyan\inst{1,2},\\ S.Mirzoyan\inst{1,3,4}, E.Poghosian\inst{1}, D.Vetrugno\inst{5}, G.Yegorian\inst{1}
}

\institute
{\inst{1} Yerevan Physics Institute and Yerevan State University, Yerevan, Armenia\\
\inst{2} School of Mathematical Sciences, Monash University, Clayton, Australia\\
\inst{3} Dipartimento di Fisica "E.R. Caianiello", Universita' degli Studi di Salerno, Fisciano, Italy\\
\inst{4} Institute of Theoretical Physics, University of Zurich, Zurich, Switzerland\\
\inst{5} Salento University and INFN, Sezione di Lecce, Lecce, Italy
}

\date{Received (\today)}

\titlerunning{CMB}

\authorrunning{V.G.Gurzadyan et al}

\abstract{
%\textbf{\boldmath
Cosmic microwave background (CMB) radiation is characterized by well-established scales, the 2.7 K temperature of the Planckian spectrum and the $10^{-5}$ amplitude of the temperature anisotropy. These features were instrumental in indicating the hot and equilibrium phases of the early history of the Universe and its large-scale isotropy, respectively. We now reveal one more intrinsic scale in CMB properties. We introduce a method developed originally by Kolmogorov, which quantifies a degree of randomness (chaos) in a set of numbers, such as measurements of the CMB temperature in a given region. Considering CMB as a composition of random and regular signals, we solve the inverse problem of recovering of their mutual fractions from the temperature sky maps. Deriving the empirical Kolmogorov's function in the Wilkinson Microwave Anisotropy Probe's maps, we obtain the fraction of the random signal to be about 20 per cent; i.e., the cosmological sky is a weakly random one. The paper is dedicated to the memory of Vladimir Arnold (1937-2010). 
}


\maketitle

The high-accuracy Planckian spectrum, the quadrupole anisotropy, and the acoustic peaks of the power spectrum (\cite{Mat,Sm,DB}) of the cosmic microwave background (CMB) appear to be carriers of valuable information on the early Universe. Among other peculiarities of the CMB signal is that it is described well by a Gaussian distribution of initial fluctuations. This is remarkable at least for two reasons. Gaussian fluctuations follow from the simplest single scalar field version of inflation. The Gaussian is also a limiting distribution, in accordance with the central limit theorem, as the sum of many independent random variables of finite variance.  CMB fitting the Gaussian certainly does not reveal its randomness, as it might arise from both random and regular distributions. 

Is it possible to reveal the fractions of random and regular signals in the CMB?  

One can deal with this problem using the Kolmogorov distribution and the stochasticity parameter as a rigorous measure of randomness (\cite{Kolm,Arnold}). Arnold applied this technique to arithmetical and geometrical progressions and to number sequences (Arnold 2008abc,2009ab), and it has also been applied to CMB temperature datasets (\cite{GK2009}). 
Now, based on the 7-year CMB data obtained by the Wilkinson Microwave Anisotropy Probe (WMAP) (\cite{J}) we have derived a general result, i.e., that the random signal yields about 0.2 fraction in the CMB total signal.  

We start by briefly outlining the approach. We consider a sequence $\{X_1,X_2,\dots,X_n\}$ of a random variable $X$ ordered in an increasing manner $X_1\le X_2\le\dots\le X_n$. Cumulative distribution function of $X$ is defined as the probability of the event $X\le x$
$$
F(x) = P\{X\le x\},
$$
and an empirical distribution function $F_n(x)$ as
\begin{eqnarray*}
F_n(x)=
%\begin{cases}
0\ , & for\, x<X_1\ ;\\
k/n\ , & for\, X_k\le x<X_{k+1},\ \ k=1,2,\dots,n-1\ ;\\
1\ , & for\, X_n\le x\ .
%\end{cases}
\end{eqnarray*}
Then the Kolmogorov's stochasticity parameter $\lambda_n$, which is also a random quantity, is defined as 
the normalized deviation of those two distribution functions
\begin{equation}\label{KSP}
\lambda_n=\sqrt{n}\ \sup_x|F_n(x)-F(x)|\ .
\end{equation}
Here, the multiplier $\sqrt n$ appears to account for the deviation expected for $n$ independent
observations of a genuinely random variable.

The remarkable result is Kolmogorov's (1933) proof (``the astonishing Kolmogorov's theorem" as Arnold (2008b) called it), concerning the asymptotic behavior of the distribution of $\lambda$ for any continuous $F$
$$
\lim_{n\to\infty}P\{\lambda_n\le\lambda\}=\Phi(\lambda)\ ,
$$
where the limiting distribution for any real $\lambda>0$ is 
\begin{equation}
\Phi(\lambda)=\sum_{k=-\infty}^{+\infty}\ (-1)^k\ e^{-2k^2\lambda^2}, \,\, \Phi(0)=0, \,   \label{Phi}
\end{equation}
the convergence is uniform, and $\Phi$, the Kolmogorov's distribution, is independent of the distribution $F$ of the initial random variable $X$.
The interval of probable values of $\lambda$ yields $0.3\le\lambda_n\le 2.4$ (\cite{Kolm}).  This universality of Kolmogorov's distribution marks it as a measure of a stochasticity degree of datasets (Arnold 2008abc,2009ab).

This objective measure enables one to consider the composition of signals of various randomness; e.g., the behavior of $\lambda_n$ and $\Phi$ was studied at numerical simulations of sequences (\cite{mod,mod1}). 

Within this remarkable system, an approach to apply is the CMB dataset, just because its cumulative distribution function happens to be known, and it is close to Gaussian! Then for Gaussian distribution function $F$ and via a properly developed strategy (for algorithmic and numerical details see Gurzadyan and Kocharyan 2008; Gurzadyan et al 2009) one can obtain the distribution of $\Phi$ for CMB, the Kolmogorov map for WMAP's temperature data, as shown in Fig.1. Not only is the non-cosmological signal, the Galactic disk, clearly outlined, which was not at all trivial {\it a priori}, but also a non-Gaussian feature such as the Cold Spot also appears to be outlined by a specific behavior by the function $\Phi$ that is  compatible with its void nature (Gurzadyan and Kocharyan 2008,2009; Gurzadyan et al 2005,2009). The Gaussian nature of CMB is characterized by certain correlations studied by traditional methods, i.e. the harmonic analysis, correlations functions, and power spectrum, which were particularly informative due to the structures of the acoustic peaks. Application of Kolmogorov's approach will complement those studies by quantifying the cumulative degree of correlations in the CMB signal.
Kolmogorov's parameter is also efficient in detecting point sources, radio and Fermi/LAT gamma-ray, and other anomalies in CMB maps (\cite{G2010,G_plane}); i.e., this method enables the separation of the cosmological signal from non-cosmological ones. 

In the present paper, for the first time, we undertake a novel analysis of the cosmological CMB signal revealed by the $\Phi$ distribution.  First, we obtain the Kolmogorov distribution $\Phi(\lambda)$ for different regions of sky and WMAP's 7-year W-band data (\cite{J}) in HEALPix representation (\cite{HP}). The procedure was as follows. Circles of radii $1^{\circ}.5$ were considered to have enough pixels to estimate the mean $\Phi(\lambda)$, given that the $\lambda$ itself is a random parameter. The variance of the Gaussian then was estimated for each circle region containing around on average 540 pixels. Then the resulting frequency distribution vs $\Phi$ was obtained, and it appears to be decaying,  corresponding to 10000 2D-balls randomly distributed over the sky with the consequent elimination of belts of coordinates $|b|< 20^{\circ}$,  $|b|< 30^{\circ}$,  $|b|< 40^{\circ}$, in order to evaluate the role of the contamination of the Galactic disk. 
Figure 2 shows the plot for $|b|< 30^{\circ}$. The 2D-balls obviously also cover the cold spot regions studied in (\cite{J,GK2009}).
The point sources revealed in CMB maps (\cite{G2010}) are also covered; however, they cannot contribute to the distribution of $\Phi$, not only because of their small quantity but also mainly because of their point-like feature, i.e. small number pixelization.    

\begin{figure}
\begin{center}
\centerline{\epsfig{file=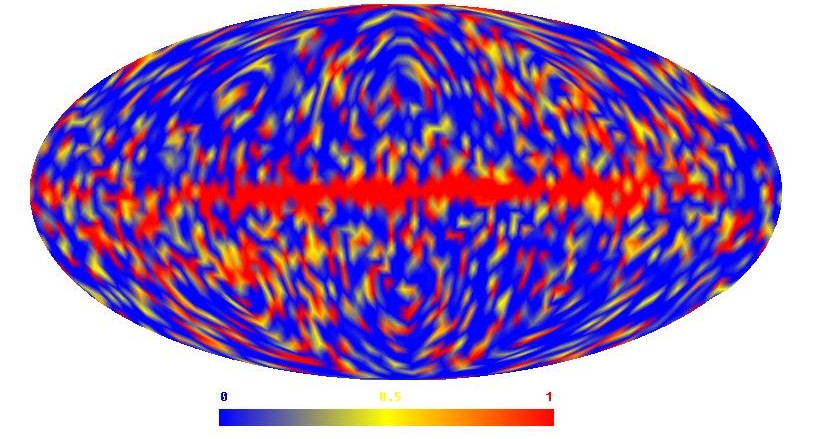,width=0.5\textwidth}} %\vspace*{2pt}
\end{center}
%\par
\caption{The Kolmogorov map of cosmic microwave background radiation, i.e. the distribution of Kolmogorov's function $\Phi$ over the sky for the Wilkinson Microwave Anisotropy Probe's 7-year W-band temperature dataset.}
\label{fig:zn_phi}
\end{figure}

\begin{figure}
\begin{center}
\centerline{\epsfig{file=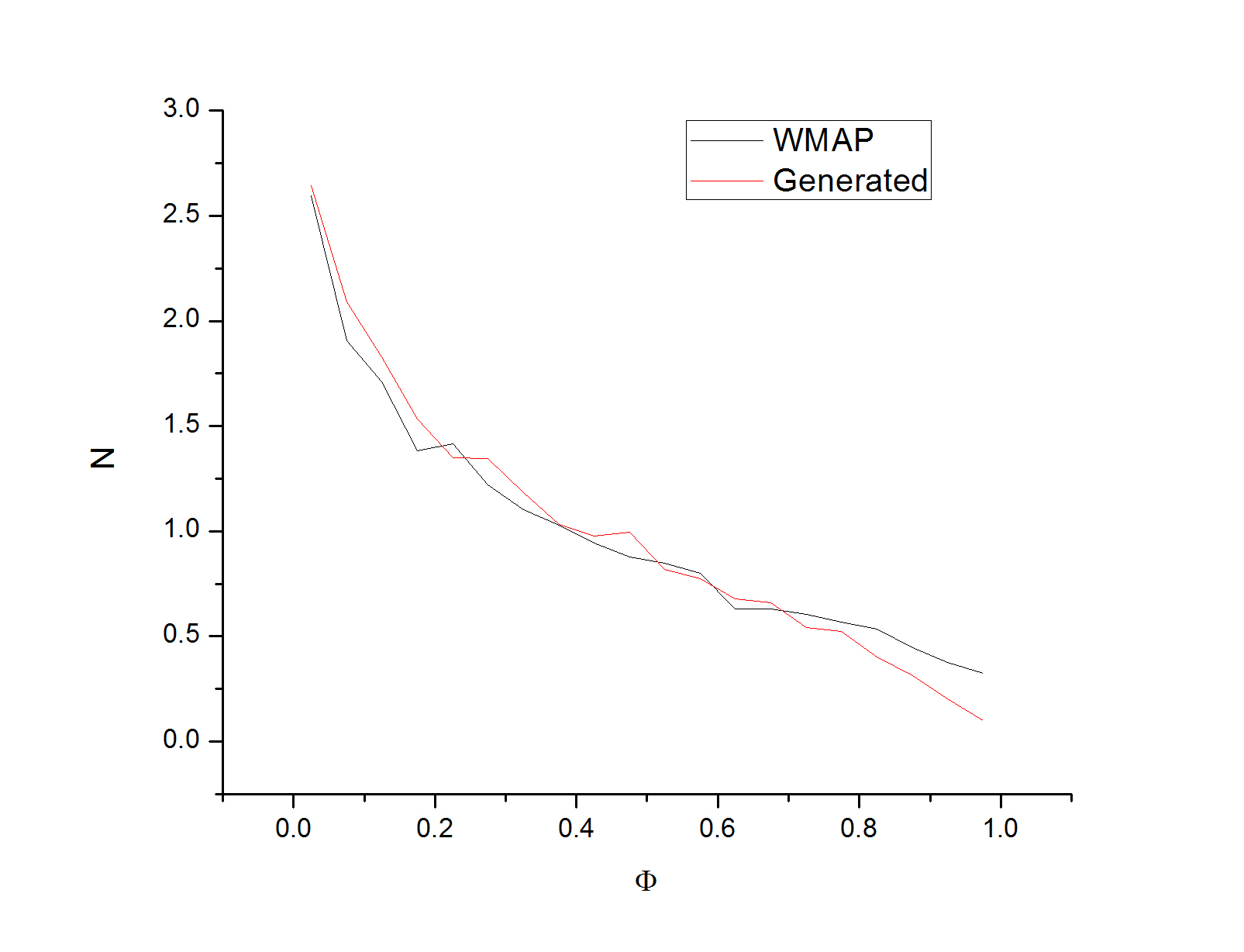,width=0.5\textwidth}} %\vspace*{2pt}
\end{center}
%\par
\caption{The empirical frequency distribution of Kolmogorov's function $\Phi$ in the CMB map with the extracted Galactic disk region $|b|< 30^{\circ}$ and the generated signal $z_n$ for the fraction of the random sequence $\alpha=0.19$.}
\label{fig:zn_phi}
\end{figure}

\begin{figure}
\begin{center}
\centerline{\epsfig{file=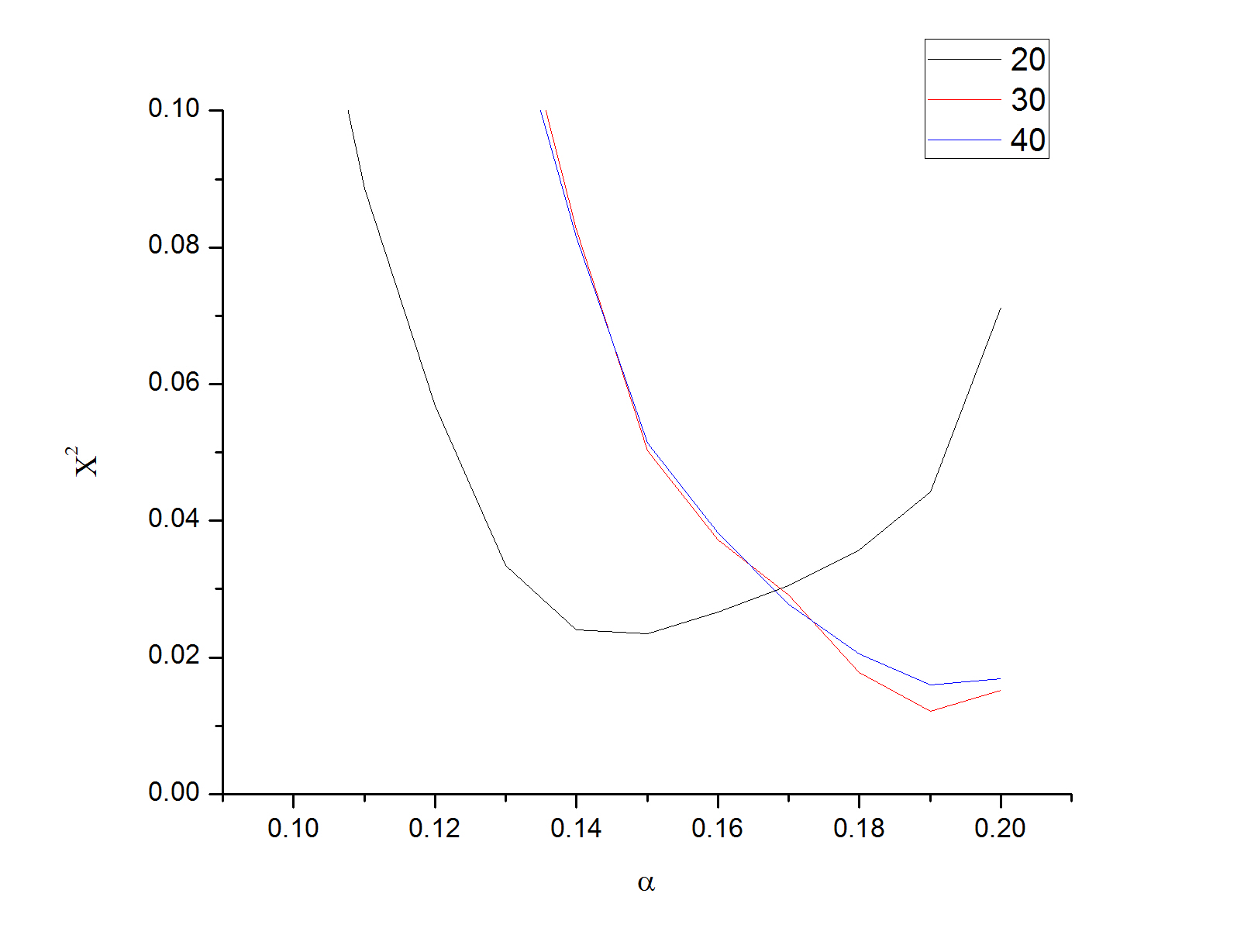,width=0.5\textwidth}} %\vspace*{2pt}
\end{center}
%\par
\caption{ The confidence level in $\chi^2$ vs the parameter $\alpha$, i.e. the fraction of the random component, for the WMAP7 CMB maps with extracted Galactic disk regions: $|b|< 20^{\circ}, 30^{\circ}, 40^{\circ}$.}
\label{fig:zn_phi}
\end{figure}

The first conclusion drawn from Figure 2 is expected. The CMB signal is certainly not purely random Gaussian; otherwise, the stochasticity parameter, hence, the function $\Phi$, would have a uniform distribution, i.e. constant, as follows from Kolmogorov theorem. 

Once Kolmogorov theorem enables estimation of the randomness $\Phi$ for a given sequence, with the empirical distribution of $\Phi$ in Figure 2, one can inquire into the inverse problem of reconstructing the initial signal to result in that distribution. 

To solve this inverse problem, we consider a regular sequence represented in the form (cf. Arnold 2008c) 
\begin{equation}
c_n = (a + h n) \mbox{ mod}(b - a),
\end{equation}
i.e. as an arithmetical progression repeated within an interval $[a,b]$, and where $h$ is an arbitrary constant. 

Since the sequence (3) corresponds to a uniform distribution, one can use von Neumann's method (\cite{vN})  
to extract a Gaussian nonrandom sequence of standard deviation $\sigma$ and mean value $m$ from it.  The latter we consider within the adopted interval $a=m-10 \sigma$ and $b=m+10 \sigma$. 

Von Neumann's method enables generating of sequences with any differential distribution  $f(x)$ and extracting of sequences with the desired distribution from a uniform one. 
The method is particularly efficient when no explicit form of $F(x)$, i.e. of the integral of $f(x)$, is available.
The procedure is as follows. First, we consider a uniform sequence $a_n$ and a distribution $f(x)$ having minimal and maximal values $y_{min}$ and $y_{max}$. Then, to extract a sub-sequence from $a_n$ one has to go through all elements of $a_n$ and, for each $a_i$, generate a random uniform number $z$ within the interval $(y_{min}, y_{max})$; if $z < f(a_i)$, one assigns $a_i$ to a new sequence and skips it otherwise. The extracted subsequence will have a distribution $f(x)$.
We used this procedure to generate random and regular sequences with Gaussian distribution, i.e. when $f(x)$ is a Gaussian. If the original sequence $a_n$ is a random one, then one gets a random Gaussian sequence, and a non-random one otherwise.

By generating many regular sequences with a Gaussian distribution of mean and standard deviations corresponding to the WMAP's CMB signal and calculating the Kolmogorov parameter for each of them, we were able to construct the empirical distribution of $\Phi$ at the best choice of the parameter $h=7.342...$.  

As the next step, the following sequence was considered, 
\begin{equation}
z_n=\alpha x_n + (1-\alpha) y_n,
\end{equation}
where $x_n$ is a random Gaussian sequence with the same $m$ and $\sigma$ as the regular von Neumann-Gaussian sequence $y_n$, i.e. the one derived by means of the von Neumann method. Thus, we have a sum of random and regular sequences with their varying proportion defined by the parameter $\alpha \in [0,1]$:  for $\alpha=0$ we have a regular sequence, while for $\alpha=1$ it is random sequence. 

It can be shown that $z_n$ also has a Gaussian distribution with mean $m$ and standard deviation $\sqrt{1-2\alpha+2\alpha^2} \sigma$. While varying $\alpha$ from $0$ to $1$ we see, as expected, the distribution of $\Phi$ gradually changing from the form in Figure 2 to a uniform one. At some value of $\alpha$, we get the closest fit evaluated in confidence levels for distribution of $\Phi$ of $z_n$ to WMAP's sequences.  Figure 3 shows the behavior of $\chi^2$ vs the $\alpha$ for the CMB maps with extracted regions $|b|< 20^{\circ}$,  $|b|< 30^{\circ}$,  $|b|< 40^{\circ}$, and as seen, the results of the last two maps are identical. We see that the minimal values of $\chi^2$, so the highest confidence levels are reached for 0.15 and 0.19 values of $\alpha$ for the $20^{\circ}$ and  $30^{\circ}$ disks, respectively.

The resulting $\alpha$  distributions are shown in Figure 2 along with WMAP's frequency plot for the CMB map with extracted Galactic disk of $|b|< 30^{\circ}$. If the small difference, a few per cent, between the values of $\alpha$ of the CMB maps with extracted $|b|< 20^{\circ}$ and $|b|< 30^{\circ}$ regions stems from the Galactic disk contamination, then one can state that the latter contributes to a slight regularization of the CMB signal.

The methodology of Kolmogorov's approach is based on studying statistical properties, namely, the degree of randomness of the signal via the stochasticity parameter and the distribution $\Phi$. As it appears, those characteristics are informative enough to distinguish one signal from another, for example, the cosmological CMB and the Galactic synchrotron one. Although the photons of such different origins can have the same temperature, they differ in their degree of randomness, hence in their correlation properties; e.g., the synchrotron is characterized by a power-law spectrum, and the CMB by a Planckian one. This difference is reflected in the statistical analysis. However, here we went further to quantify the contribution of the correlations and of chaotic components. The correlations in the CMB signal are carriers of information on various processes, from primordial, through the last scattering and reionization epochs up to the integrated Sachs-Wolfe and Sunyaev-Zeldovich effects. Revealing the degree of correlations in CMB signal, particularly, reflects the cumulative contribution and properties of those processes.         

Thus, considering CMB as a composition of signals, we showed that the random component is a minor perturbation of mostly regular signal. The model-independent character of this result comes from the universality of the Kolmogorov theorem that we used in analysing real datasets, the WMAP's 7-year CMB maps. We solved the inverse problem by numerical methods and recovered the mutual fractions of random and regular signals which would lead to the empirical Kolmogorov's distribution. 

That the CMB sky is a weakly random one and the derived 0.2 fraction of the random signal in CMB by their intrinsic statistical content may be compared with other scales of CMB, the 2.7K temperature of the Planckian spectrum, and the $10^{-5}$ amplitude of the quadrupole anisotropy, indicating the hot equilibrium history of the Universe and the isotropy of the Universe, respectively. If the Planckian spectrum and the anisotropy amplitude do reflect and do scale certain correlations in CMB signal, the obtained fraction scales the correlations in the cosmological Gaussian perturbations.

What insights  will the new scaling open to the early Universe?

\vspace{0.2in}

{\it We dedicate this paper to the memory of Vladimir Arnold (1937-2010), recalling the memorable discussions (VG,AAK) with him.}

\end{document}